\documentclass[10pt,conference]{IEEEtran}
\usepackage{xspace}
\usepackage{multirow}
\usepackage{multicol}
\usepackage{graphicx}
\usepackage{tikz}
\usepackage{xcolor}
\usepackage{url}
\usepackage{hyperref}
\usetikzlibrary{positioning, fit, scopes}
\usepackage{pgfplots}
\pgfplotsset{compat=1.15}
\usepackage[para,online,flushleft]{threeparttable}

\newcommand{\TOOL}{\texttt{class-bot}\xspace}
\newcommand{\drca}{\textit{developer recommendation choice architectures}\xspace}

\begin{document}

\title{Nudging Students Toward Better Software Engineering Behaviors}

\author{\IEEEauthorblockN{Chris Brown, Chris Parnin}
\IEEEauthorblockA{\textit{Department of Computer Science} \\
\textit{North Carolina State University}\\
Raleigh, NC, USA \\
dcbrow10@ncsu.edu, cjparnin@ncsu.edu}
}

\maketitle

\begin{abstract}
Student experiences in large undergraduate Computer Science courses are increasingly impacted by automated systems. Bots, or agents of software automation, are useful for efficiently grading and generating feedback. Current efforts at automation in CS education focus on supporting instructional tasks, but do not address student struggles due to poor behaviors, such as procrastination. In this paper, we explore using bots to improve the software engineering behaviors of students using \drca, a framework incorporating behavioral science concepts in recommendations to improve the actions of programmers. We implemented this framework in \TOOL, a novel system designed to \textit{nudge} students to make better choices while working on programming assignments. This work presents a preliminary evaluation integrating this tool in an introductory programming course. Our results show that \TOOL is beneficial for improving student development behaviors increasing code quality and productivity.
\end{abstract}

\section{Introduction}

Enrollment in undergraduate Computer Science (CS) courses is growing rapidly, and these classes are quickly evolving to accommodate the significant increase of students~\cite{Kay98Intro}. To handle the large influx of students, researchers and practitioners have developed a variety of bots to automate instructional tasks. For instance, tools such as CoderAssist~\cite{Kaleeswaran2016CoderAssist}, Web-CAT~\cite{Edwards08WebCAT}, and AutoGradr\footnote{https://autogradr.com}  are useful for grading programs and providing feedback on students' code. %For example, Kaleeswaran and colleagues show CoderAssist automatically grades C programs, provides useful feedback to students on their code, and grades projects in an average of 1.6 seconds using \textit{semi-supervised verified feedback generation}~\cite{Kaleeswaran2016CoderAssist}. 
Furthermore, Wilcox shows using automated tools to facilitate student learning in programming courses is valuable for improving performance, increasing submissions, and saving instructional time~\cite{Wilcox2015Automation}.

Despite increasing enrollment in CS courses, research shows the dropout rate in programming classes is also increasing rapidly, especially among first and second year students~\cite{beaubouef2005high}. Studies suggest the inability to maintain students in CS will lead to a ``crisis'' in the software industry with necessary computing-related jobs going unfilled.\footnote{https://www.itexico.com/blog/how-to-handle-the-crisis-of-software-developer-shortage-in-the-u.s} Consequently, underrepresented minorities and female students disproportionately account for those underperforming in early programming courses and dropping out of the CS major~\cite{Payton20NC, Baer20Gender}, leading to a lack of diversity in industry.\footnote{https://news.gallup.com/reports/196331/diversity-gaps-computer-science.aspx}

Automated systems are beneficial for providing feedback efficiently, however they lack the ability to support students. The goal of this work is to explore using bots to retain students in CS. While not solely due to a lack of effort, Beaubouef and Mason suggest one reason for high withdrawal in programming courses is poor behavior on coding assignments~\cite{beaubouef2005high}. Decision-making is an important skill in software engineering,\footnote{https://hackernoon.com/decision-making-the-most-undervalued-skill-in-software-engineering-f9b8e5835ca6/} however students in programming courses frequently make poor choices and adopt bad programming behaviors when writing code for projects. For example, students often ignore software development processes, which leads to increased frustration, lower grades, and eventually abandoning CS~\cite{beaubouef2005high}. %Furthermore, these behaviors persist among professional software engineers who also often underestimate the time and effort required to complete coding tasks~\cite{Boehm1984SEEcon}, inadequately test software~\cite{Whittaker00Testing}, and lack sufficient documentation in their code~\cite{Briand03Documentation}. Similarly in software engineering, bots are useful for automating programming tasks, however developers often find automated recommendations from these systems ineffective and intrusive~\cite{SorryToBotherYou}.

Our prior work proposes using \textit{nudge theory} to improve the reception of automated recommendations to developers from bots~\cite{SorryToBotherYou,SorryToBotherYou2}. Nudge theory is a behavioral science concept for improving human behavior by influencing the environment surrounding decisions, or choice architecture, without providing incentives or banning alternative choices~\cite{sunstein2008nudge}. We introduced \drca, a novel framework for designing recommender bots to nudge software engineers towards better development practices~\cite{SorryToBotherYou2}. In this work, we explore implementing this framework in an innovative system to improve the software engineering behaviors of students on programming projects. % To discover the impact of nudges in CS education, we seek to answer the following research questions:

% \begin{itemize}
%   \item[\textbf{RQ1}] How do nudges impact the quality of student projects?
%   \item[\textbf{RQ2}] How do nudges influence student productivity?
% \end{itemize}

To discover the impact of this approach on student behavior, we performed a study implementing \TOOL, a system that utilizes \drca to recommend beneficial software engineering behaviors to students on projects for a university-level introductory programming course. We evaluated the effectiveness of this system by examining the code quality of projects and productivity of students. Our results suggest automated nudges with \drca improved student behavior by increasing performance on assignments and encouraging students to make more substantial changes to their code and start work earlier. The contributions of this research are a novel bot to recommend software engineering practices to students and a preliminary evaluation examining the impact of \drca on improving the software engineering behaviors of programmers.

\section{Background}

%This research is based on prior work exploring behavioral science as well as tools for Computer Science education.

\subsection{Nudge Theory}

Behavioral science research suggests nudges are effective for improving human decision-making. For example, studies show placing healthy options near the front of a cafeteria encourages students to purchase and consume healthier options~\cite{Hanks2012Lunchroom}. In this case, students are not rewarded for making healthy decisions nor banned choosing from junk food. Furthermore, integrating nudge theory in digital choice environments is known as \textit{digital nudging}~\cite{weinmann2016digitalnudging}. Prior work suggests digital nudges can encourage better privacy and security decisions of users online~\cite{Acquisti2017Nudges} and improve student learning and motivation~\cite{Rughini2013DigitalBadge}. We aim to explore using digital nudges to improve the programming behaviors of students on coding assignments to improve their code quality, productivity, and ultimately increase retention among early CS students.

Nudges are useful for improving human behavior because of their ability to influence the context and environment surrounding decisions, or \textit{choice architecture}~\cite{thaler2013choice}. Brown and colleagues propose \textit{developer recommendation choice architectures}, a framework that applies concepts from nudge theory into automated bots to create effective developer recommendations~\cite{SorryToBotherYou2}. We incorporate this framework into \TOOL to make recommendations encouraging students to adopt better programming behaviors while working on projects. Our goal is to discover the impact of \drca on the decision-making and behavior of students by analyzing code quality and development productivity on programming assignments in introductory CS classes.

\subsection{Computer Science Education}

CS education researchers have evaluated a variety of systems useful for automating instructional tasks. For example, Kaleeswaran et al. found CoderAssist grades C programs and provides feedback to students on their code in an average of 1.6 seconds~\cite{Kaleeswaran2016CoderAssist}. Prior work also outlines strategies for incorporating automated grading tools on coding assignments~\cite{Wilcox16AutoGradingStrategies}. Moreover, Heckman suggests incorporating real development tools in courses can improve software engineering skills~\cite{Heckman18Tools} while Hu shows integrating GitHub pull request bots on project repositories provides fast and useful feedback to students~\cite{hu2019feedback}. In this work, we seek to use bots as a means to enhance development behaviors in introductory programming courses.

Researchers have also explored improving student behavior in CS courses. For example, Edwards and colleagues summarize effective and ineffective student behaviors on programming assignments~\cite{Edwards09Behaviors}. Studies also show that non-automated approaches, such as in-class labs~\cite{Heckman15Labs}, agile practices~\cite{Schneider2020Agile}, pair learning~\cite{williams2000effects}, and other instructional techniques improve student behavior and learning in undergraduate CS courses. Additionally, prediction models~\cite{Niki19Predictive} and automated tools such as DevEventTracker~\cite{DevEventTracker} are useful for observing student programming habits to predict bad behaviors, such as procrastination. This research explores using automated notifications from bots to improve the decision-making of students while developing code on programming assignments.

\section{CLASS-BOT}

\begin{figure}
\centering
	\includegraphics[width=0.5\textwidth]{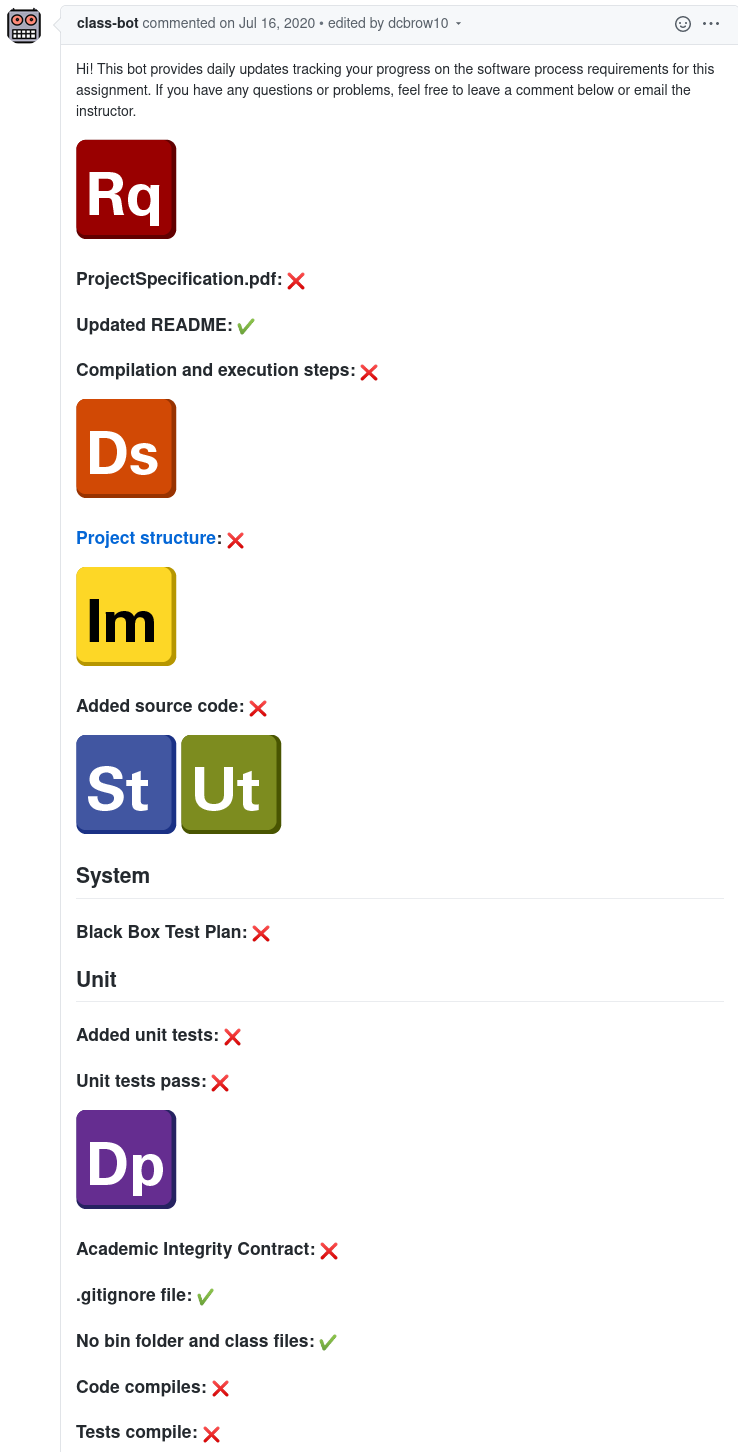}
	\caption{Example \TOOL recommendation}	
	\label{fig:bot} 
\end{figure}

The \TOOL system nudges students by automatically generating and updating GitHub issues on project repositories (see Figure~\ref{fig:bot}). We implemented our bot using the GitHub issue tracker because this system is useful for managing bugs, enhancements, and feedback on repositories,\footnote{https://help.github.com/en/articles/about-issues} and research suggests they are useful for making recommendations to developers~\cite{bissyande2013issues}. To improve student programming behaviors, \TOOL encouraged them to complete the \textit{software development process}, or set of activities necessary to program software application. Beaubouef and Mason suggest students' failure to follow development processes factors into the high attrition rate in early programming courses, noting the typical student method ``\textit{minimally includes the processes of analysis, design, coding, testing, and documentation...Unsuccessful students often want to skip analysis and design and begin typing in code immediately.}''~\cite[p.~105]{beaubouef2005high}

Each \TOOL issue contains sections for each software development process phase (i.e. \includegraphics[height=1em]{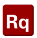} in Figure~\ref{fig:bot} represents the Requirements phase) and listing relevant project rubric items for that phase. Specific tasks differed based on the assignment. In general: Requirements (Rq) focuses on understanding project specifications and functionality; Design (Ds) relates to project structure; Implementation (Im) centers on the code; Testing ensures students added passing unit tests (Ut) and functional test cases (St); and Deployment (Dp) verifies the repository is ready for submission and grading based on submission instructions. We compared our system to a baseline approach using an online rubric with a similar organization.

The \TOOL issue updates fit the definition of a nudge because they do not provide rewards for completing tasks nor prevent students from ignoring items. To improve student programming behaviors, we designed \TOOL using the three principles of \drca: \textit{Actionability}, \textit{Feedback}, and \textit{Locality}~\cite{SorryToBotherYou2}. Here, we explain how \TOOL incorporates each principle to recommend better development process behaviors to students.

\paragraph{Actionability}

Actionability involves reducing user effort by automating tasks to facilitate the adoption of useful behaviors. We implemented \TOOL to incorporate actionability by automatically analyzing repositories to determine if development process tasks are completed and update GitHub issues based on code changes. However, the baseline approach requires students to manually seek information online.

\paragraph{Feedback}

Feedback refers to the clarity of information provided to users. Our system provides a simple feedback mechanism, displaying a red x (\includegraphics[height=1em]{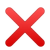}) if the requirements for an item are not met and a green check mark (\includegraphics[height=1em]{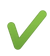}) if the task is completed. For instance, in the Deployment phase of the \TOOL example in Figure~\ref{fig:bot}, the repository contains a .gitignore file but the code does not compile. With the the baseline approach, students are forced to check if project expectations are met themselves.

\paragraph{Locality}

Locality focuses on the setting of a recommendation, or when and where automated interventions occur. For \textit{spatial} locality, \TOOL recommendations are located in an issue on the repository situated with the project. For \textit{temporal} locality, or notification timing, \TOOL provides automated daily updates on software development processes based on students' code contributions. On the other hand, students are forced to search for information in an ad hoc manner at a separate location from their repository using the online rubric.

\section{Methodology}

%To evaluate \drca, we conducted a preliminary study observing the impact of \TOOL on student behavior.

\subsection{Experiment Design}

\subsubsection{Participants} We integrated \TOOL in an introductory Java programming course. All participants were undergraduate students with varying majors and levels of programming experience. For consistency in our data, we eliminated students who eventually dropped the class. Overall, we analyzed the behavior of 35 out of the initial 42 enrolled students. We use five phases to define the software development process: \textit{Requirements}, \textit{Design}, \textit{Implementation}, \textit{Test}, and \textit{Deployment}, as presented to students in the course curriculum.

\subsubsection{Projects}

The course consisted of seven programming assignments, six projects and a final comprehensive exercise. Projects 3-5 made up the control group to avoid beginning assignments, and \TOOL was introduced on the final two assignments. All projects were submitted to individual GitHub repositories. In total, we analyzed a 151 project repositories.

\subsection{Data Collection}

To determine the impact of \TOOL on student behavior, we mined project repositories to observe the quality of their work and development productivity. Data analysis and collection lasted over the course of approximately six weeks.

\subsubsection{Quality}

We evaluated quality by observing the \textit{grades} and number of \textit{points deducted} for project submissions. In the course, project grades were determined using realistic industry code quality metrics such as passing unit tests, functional test cases, and Checkstyle\footnote{\url{https://checkstyle.sourceforge.io/}} static analysis tool warnings. Additionally, students who failed to meet certain project requirements had additional points subtracted from their grade. For instance, submitting an assignment within 24 hours after the deadline resulted in a -10\% late penalty. We aim to discover if \TOOL impacts student behavior by improving the quality of their projects. %, .% For coding projects, the assignment grade indicates the overall quality of the project. % Research suggests poor project management skills result in low grades on programming assignments~\cite{beaubouef2005high}. Conversely, software engineering teams with higher software process maturity produce higher quality programs~\cite{Clark97theeffects}. 

\subsubsection{Productivity}

We measured productivity by observing the total \textit{number of commits}, \textit{code churn}, time until the \textit{first commit}, and time of the \textit{last commit}. GitHub commits record specific changes to project files,\footnote{https://docs.github.com/en/desktop/contributing-and-collaborating-using-github-desktop/committing-and-reviewing-changes-to-your-project\#about-commits} and prior work explores using repository commits to predict student performance~\cite{sprint2019mining} and gamify contributions to projects~\cite{Singer12Race}. Research also suggests code churn, or the number of lines added, deleted, or modified in commits, is useful for measuring developer effort and code change impact~\cite{Munson98Churn}. Moreover, prior work shows students who start assignments earlier receive significantly higher grades while those who procrastinate often perform worse~\cite{Edwards09Behaviors}. To further study productivity, we measured the time until the first commit on repositories and the amount of time between the last commit and the assignment deadline. 

We use these quality and productivity metrics to investigate the impact of bots incorporating \drca on the software engineering behaviors of students working on programming projects.
 
%\subsubsection{Survey}
 
%Finally, we surveyed students to collect feedback on their experience with \TOOL. Out of the 35 eligible students, we received 9 responses (25.7\% response rate). We analyzed open-ended responses to derive themes and gain insight into improving automated tools for recommending beneficial behaviors to students. % The survey consisted of a 5-point Likert scale to ask students how useful they found nudges from \TOOL and presented open-ended and free response questions for students to provide feedback on the system.

\begin{table}[tbh]
\centering
\caption{Quality Results}
\begin{threeparttable}
\begin{tabular}{ |l|l|r|r|c| } \hline
  \textbf{} & \textbf{Nudge?} & \textbf{Mean} & \textbf{Median} & \textbf{\textit{p-value}} \\ \hline
 \multirow{2}{*}{\textbf{Grade***}} & No & 74.29 &  87.66 & - \\
 & Yes & 76.89 & 95 &  \textbf{\em 0.0097***}  \\ \hline
 \multirow{2}{*}{\textbf{Deductions}} & No & -20.71 & -5 & - \\
 & Yes & -9.43 & 0 &  0.0672 \\ \hline
\end{tabular}
\begin{tablenotes}
% \centering
Quality metrics for projects with and without \TOOL \\ \textbf{***} denotes statistically significant results (\textit{p-value} $< 0.05$)
\end{tablenotes} 
\end{threeparttable}
\label{tab:quality}
\end{table}

\begin{table}[tbh]
\centering
\caption{Productivity Results}
\begin{threeparttable}
\begin{tabular}{ |l|l|r|r|c| } \hline
  \textbf{} & \textbf{Nudge?} & \textbf{Mean} & \textbf{Median} & \textbf{\textit{p-value}} \\ \hline
 \multirow{2}{*}{\textbf{Commits}} & No & 9.84 & 7 & - \\
 & Yes & 12.64 & 9 & 0.1646 \\ \hline
 \multirow{2}{*}{\textbf{Code Churn***}} & No &  205.03 & 4 & - \\
 & Yes & 1101.57 & 11 & \textbf{\em 0.0348***} \\ \hline
 \multirow{2}{*}{}\textbf{First Commit***} & No &  8.32 & 7.41 & - \\
 \textbf{(days)} & Yes & 1.99 & 5.94 & \textbf{\em $<$ 0.0001***} \\ \hline
  \multirow{2}{*}{}\textbf{Last Commit} & No &  -21.72 & -1.60 & - \\
 \textbf{(hours)} & Yes & -9.67 & -2.47 &  0.7909 \\ \hline
\end{tabular}
\begin{tablenotes}
% \centering
Productivity metrics for projects with and without \TOOL \\ \textbf{***} denotes statistically significant results (\textit{p-value} $< 0.05$)
\end{tablenotes} 
\end{threeparttable}
\label{tab:productivity}
\end{table}

\section{Results}

The Mann-Whitney-Wilcoxon test ($\alpha$ = .05) was used to compare project quality and student productivity metrics for assignments with and without nudges from \TOOL.

\subsection{Project Quality}

To observe the impact of \TOOL on code quality, we analyzed the grade and points deducted on student assignments. These findings are presented in Table~\ref{tab:quality}. We discovered students received significantly higher scores on projects with automated nudges, indicating notifications from our system improved the quality of programming assignments. Additionally, while there was not a significant difference, we noticed projects without automated nudges lost an average of 11 more points than those with \TOOL interventions.

\subsection{Student Productivity}

To discover the impact of \TOOL on student productivity, we analyzed total commits, code churn, first commit time, and last commit time. Our results in Table~\ref{tab:productivity} show nudges improved the productivity of students by significantly increasing the number of lines of code modified in commits and encouraging students to start development on programming assignments earlier. Additionally, we found that on average projects with automated nudges had more commits to the repositories and projects were submitted earlier with \TOOL notifications.

% \subsection{Survey}
% We surveyed students to collect feedback on their experience with \TOOL. Out of the 35 eligible students, we received 9 responses (25.7\% response rate). Figure~\ref{fig:survey} shows most students found the notifications from \TOOL at least moderately useful. We analyzed open-ended responses to derive themes and gain insight into improving automated tools for recommending beneficial behaviors to students, which are presented in the Discussion.
 % Students praised notifications from \TOOL in their feedback noting they ``\textit{liked the class-bot updates}'' (P1) and used it to ``\textit{make sure everything was running smoothly}'' (P6). 

% \begin{figure}[htb]
% \begin{tikzpicture}
% \begin{axis}[
%     xbar stacked,
%     ytick=data,
%     axis y line*=none,
%     axis x line*=bottom,
%     tick label style={font=\footnotesize},
%     legend style={font=\footnotesize},
%     label style={font=\footnotesize},
%     xtick={0,2,4,6,8},
%     width=.4\textwidth,
%     bar width=6mm,
%     xlabel= Number of Participants,
%     yticklabels={Not at All Useful, Somewhat Useful, Moderately Useful, Useful, Very Useful},
%     xmin=0,
%     xmax=8,
%     area legend,
%     y=8mm,
%     enlarge y limits={abs=0.625},
% ]

% \addplot[fill=black] coordinates
% {(0,0) (1,1) (5,2) (2,3) (1,4)};
% \end{axis}  
% \end{tikzpicture}
% \caption{Survey Responses on the Usefulness of \TOOL}
% \label{fig:survey}
% \end{figure}

\section{Discussion}

Our preliminary results suggest bots incorporating \drca can improve development behavior. By encouraging students to follow software engineering processes, we found \TOOL improved project quality and student productivity by boosting grades, increasing code churn, and preventing procrastination. We believe these automated nudges are a step towards reducing the attrition rate in CS. Furthermore, these results provide implications for software engineers in industry where developers also ignore development processes and underestimate the time and effort required to complete coding tasks~\cite{Boehm1984SEEcon}. Despite the advantages of bots for automating programming tasks, developers find recommendations from bots ineffective and intrusive~\cite{SorryToBotherYou}. To gain insight for improving bots to effectively recommend beneficial programming behaviors, we analyzed feedback from students and found increased project validation as well as more frequent updates as reported improvements for the \TOOL system.

\paragraph{Validation Frequency}

Even though they started assignments approximately six days earlier with \TOOL, students still reported waiting until the end of the project to validate their work met project expectations. For example, one student noted ``\textit{I didn't really check them until the final day}'' and another mentioned ``\textit{I checked it once at the end to make sure everything was correct but thats it}''. This indicates a need to encourage students to validate their code more frequently, which research shows improves code quality~\cite{Wallace89Verification}. \TOOL provides automated updates for students to passively check their progress, however future systems can be improved by incorporating interactive mechanisms, such as checklists~\cite{Garcia08Ten}, to encourage students to validate their code more frequently during the development of their projects.

% Software practitioners also suggest software validation and verification are in CS education and student programming assignments~\cite{McMillan99Emph}. Verifying and validating software are time-consuming, costly, and challenging, and Garcia et al. propose solutions such as improved training, guidelines, and checklists to overcome challenges in these processes~\cite{Garcia08Ten}.

\paragraph{Update Frequency}

Students also desired more frequent updates from \TOOL, saying they ``\textit{would like it more if it could be updated more often, or maybe later in the day}'' and ``\textit{the class bot didn't update frequently enough}''. While \TOOL incorporated \textit{temporal locality} by updating issues daily, students desired more frequent feedback to support them in their work. Prior work in software engineering shows that presenting recommendations to developers at ``diff time'' during code reviews encouraged developers to fix more bugs compared to less frequent notifications, i.e. after overnight builds~\cite{Distefano2019Facebook}. Based on this feedback for \TOOL, future systems can be improved by providing frequent notifications based on student actions, such as immediately after commits, to provide convenient and persistent recommendations for useful software engineering behaviors.

\section{Conclusion}

This work explores using bots to improve student software engineering behaviors in introductory programming courses. We introduce \TOOL, a novel system that incorporates \drca to create automated recommendations with clear feedback in a convenient setting to encourage students to adhere to software development process phases. Our results show this system significantly improved code quality and student productivity by raising grades, increasing coding activity, and encouraging students to start earlier. We conclude by providing implications for improving \TOOL and designing future automated systems to recommend beneficial behaviors to programmers.

\bibliographystyle{abbrv}
\bibliography{main}

\end{document}